# Псориаз как следствие включения β-стрептококков в микробиоценоз кишечника с повышенной проницаемостью (концепция патогенеза).

*Н.Г. Короткий [1], М.Ю. Песляк [2]*

# Psoriasis as consequence of incorporation of β-streptococci in the microbiocenosis of highly permeable intestines (a pathogenic concept)

**Korotkii NG, Peslyak MY.**

[1] Российский государственный медицинский университет
[2] Издательство "Кудиц-образ"; интернет-сайт www.psora.df.ru

**Представлен** аналитический обзор результатов новейших исследований патогенеза псориаза. Систематизированы данные о взаимосвязи присутствия в организме β-стрептококков и кожной иммунной реакции. Проанализированы исследования патологии ЖКТ больных псориазом. Предложена гипотеза о первичности псориатической гастроинтестинопатии по отношению к кожным проявлениям. Сформулирована модель патогенеза хронического псориаза (chronic plaque psoriasis), основанная на предположении о решающей роли двух псорафакторов - повышенной проницаемости кишечных стенок для определенных белков и включении β-стрептококков в микробиоценоз слизистой кишечника. Описана практика лечения, подтверждающая эту модель.

**Ключевые слова:** псориаз, патогенез, кожная иммунная система, кишечная проницаемость, микробиоценоз кишечника, β-стрептококки

**A review** of recent investigations into pathogenesis of psoriasis summarizing data on the relationship between the infection by β-streptococci and the immune reaction of the skin. Results of the studies on gastrointestinal pathology in psoriatic patients are presented. The authors propose a hypothesis that advocates the primary origin of psoriatic gastrointestinal pathology and secondary nature of skin responses. A chronic plaque psoriasis model is developed based on the assumption on the key role of two psorafactors, i.e. enhanced permeability of the intestinal wall for certain proteins and incorporation of β-streptococci in the microbiocenosis of intestinal mucosa. The validity of the model in confirmed by the results of practical clinical work.

Настоящая статья представляет собой обзорно-аналитическое исследование взаимосвязи нарушений в ЖКТ и кожных проявлений псориаза. Исходным материалом послужили в первую очередь результаты, изложенные в книгах [1,2,3,4], увидевших свет в новом XXI веке. И в более ранних работах уже были установлены органические отклонения при псориазе, например, в дуоденальной слизистой оболочке, в частности увеличение числа тучных клеток и эозинофилов, либо увеличение числа дуоденальных внутриэпителиальных лимфоцитов [2,5]. Рядом авторов была установлена взаимосвязь между псориазом и хроническим еюнитом (с повышенной проницаемостью тонкого кишечника и сглаживанием его слизистой ) [6,7,8]. Однако систематические исследования в этой области были проведены только в последние годы.

В работе [2] были обследованы 45 человек, имеющих только псориаз. У всех пациентов этой группы обнаружено нарушение проницаемости кишечных стенок для углеводов и жиров, коррелирующее со степенью тяжести псориаза и длительностью заболевания. Гастроскопия, выполненная для части пациентов (20 чел.), показала у всех диффузный дуоденит.

В книге [1,стр.64] обобщены результаты обследования ЖКТ 250 пациентов, имеющих только псориаз. У всех пациентов выявлена патология верхних отделов пищеварительного тракта. Степень уменьшения высоты эпителиального пласта коррелировала с прогрессирующей стадией псориаза и длительностью заболевания. Исследование биоптатов толстой кишки 20 пациентов показало наличие "*дегенеративно-дистрофических изменений эпителиального компартамента, затрагивающих в первую очередь всасывающий аппарат колоноцитов*". Авторы книги делают вывод, что у больных псориазом "*выраженные дегенеративно-дистрофические изменения клеточных популяций покровного и железистого гастроинтестинального эпителия с деструкцией функционально ведущих цитоплазматических органелл эпителиоцитов, приводят в нарушению процессов секреции и всасывания*" [1, стр.154]. Они определяют эти изменения как "*псориатическую гастроинтестинопатию*" и предполагают ее вторичность по отношению к кожным проявлениям псориаза.

В работе [7] отмечена взаимосвязь между псориазом и микрофлорой кишечника, в работах [9,10] одной из наиболее вероятных причин повышенной кишечной проницаемости названа смена микробиоценоза.

Известна роль стрептококковой фокальной инфекции для возникновения каплевидного и пятнистого псориаза [3, гл. 5]. Там же и в ряде последующих исследований [11,12,13,14] было показано, что основными β-стрептококковыми антигенами (всюду далее BS-антигенами), провоцирующими и поддерживающими хронический псориаз являются BSP-антигены (β-Streptococci Proteins) - стрептококковые оболочечные и мембранные белки, являющиеся продуктами распада BS. В работе [13] показано, что имеет место усиленная кожная иммунная реакция на стрептококковые белки клеточной оболочки с массой 20-50 kDa. Эти исследования, ведущиеся под руководством профессора Лайонела Фрая (Lionel Fry) в Великобритании, направлены на точное определение BS-антигенов, провоцирующих начало и поддержку хронического псориатического процесса.

В главе 8 книги [3] сформулированы три варианта патогенеза псориаза. Первый основан на присутствии в коже BS-антигенов, второй - на сочетании присутствия BS-антигенов и перекрестно-реактивной аутоантигенной детерминаты (например кератинов) и третий - исключительно на присутствии аутоантигенной детерминанты.

Результаты работы [15] при длительном течении псориаза (более 10 лет) свидетельствуют в пользу аутоантигенного варианта: аутоантитела к фибробластам (АФ) обнаружены у 62% больных, однако при длительности заболевания менее 1 года АФ обнаружены лишь у 7%.

Для инициации и развития псориатического процесса автор книги [3] (и авторы данной статьи) считает BS-антигенный вариант наиболее вероятным. Однако остается открытым вопрос о причинах присутствия BS-антигенов в коже после прекращения фокальной BS-инфекции или в тех случаях, когда такой инфекции не было. Попыткой ответить на этот вопрос будет предложенная далее модель патогенеза псориаза. Для ее формулировки понадобятся результаты исследований проницаемости кишечника и процессов формирования его микрофлоры.

## 1. Кишечник. Строение и заселение слизистой, проницаемость

Складки тонкой кишки изнутри выстланы ворсинками, содержащими сеть кровеносных и лимфатических сосудов. Главными клетками ворсинок являются энтероциты, на обращенном в сторону кишечного просвета участке покрытые микроворсинками, увеличивающими всасывательную поверхность кишечника до 350 кв. м. В норме цикл жизни энтероцитов составляет 3-7 дней. В тонком кишечнике суточные объемы всасывания составляют около 8,5 л, в толстом кишечнике - 0,5 л.

Дети рождаются со стерильным кишечником, но уже при прохождении родовых путей и последующем контакте с матерью начинает формироваться микрофлора толстого кишечника (окончательный баланс достигается к 3 годам). Колонизация бактерий в ЖКТ осуществляется как бы по этажам. В желудке здорового человека их практически нет, верхние отделы тонкой кишки относительно свободны от бактерий (менее 1000/мл), но нижние отделы тонкой кишки и, особенно, толстая кишка представляют собой огромный резервуар бактерий (до $10^{12}$ на мл фекальных масс). Малая часть микробов обитает в просвете кишечника, а основная масса образует колонии на стенках. Колонии покрывают эпителий кишечника многослойной биопленкой, находясь с ним в симбиозе, плотно сцепляясь с микроворсинками. Толщина и плотность этой биопленки растет по мере приближения к толстой кишке, она обеспечивает качество пристеночного пищеварения и эффективное функционирование энтероцитов и колоноцитов.

Стерильность кишечника вредна для организма. Нормальный микроэкологический баланс кишечника способствует резистентности против кишечных инфекций, усвояемости питательных веществ и, в целом, ведет к увеличению продолжительности жизни индивидуума. В список нормальных не попадают патогенные микроорганизмы, однако и остальных, присутствие которых допустимо и/или необходимо в ЖКТ, насчитывается более 400 видов. Вопрос о том, какое именно количественное и видовое сочетание микроорганизмов (свое для каждого из отделов ЖКТ) является правильным для всех, некорректен. Правильное сочетание определяется не только питанием, расой и возрастом человека, но и конкретными особенностями индивидуума, правильным является широкий спектр сочетаний, когда присутствие каждого из видов не выходит за рамки определенного, достаточно широкого, диапазона [16].

Химус движется по тонкой кишке благодаря ее перистальтической активности и преобразуется. Содержащиеся в нем вещества (в том числе микроорганизмы, продукты их жизнедеятельности и распада) из-за разницы осмотических давлений стремятся попасть через кишечные стенки в кровеносное и/или лимфатическое русло. Энтероциты, покрытые биопленкой, обеспечивают всасывание благодаря множеству биохимических процессов (отличающихся в зависимости от того, что всасывается). Различают: а) активный транспорт, б) пристеночное пищеварение (расщепление и преобразование при участии микроорганизмов, кишечного сока, ферментов, выделяемых энтероцитами), включая последующий транспорт и в) барьерную функцию. Значимые нарушения любого из этих процессов приводит к синдрому мальабсорбции, т.е. к неполному или, наоборот, чрезмерному всасыванию [17].

Микроорганизмы, составляющие биопленку, активно участвуют в процессе усвоения и всасывания химуса. Одну часть химуса они потребляют сами, другую - перерабатывают и превращают в вещества, усвоение которых макроорганизмом значительно упрощается или просто становится возможным. Фактически биопленка является частью пищеварительной системы макроорганизма. Вещества, являющиеся продуктами жизнедеятельности и распада микроорганизмов, составляющих биопленку, пополняют химус, т.е. и для них также справедливо все вышесказанное.

Для каждого из веществ химуса может быть определена норма проницаемости (всасывательной функции) кишечника – это скорость всасывания этого вещества в единицу времени при определенном его количестве, поступившем в кишечник. Если проницаемость для конкретного вещества заметно отличается от нормы (больше или меньше) – это свидетельствует о нарушениях всасывательной функции.

В статье [6] (см. раздел 4) и в монографии [2] приведены результаты исследований всасывательной функции у больных псориазом. В последней работе обследованы 45 пациентов с псориазом и у всех обнаружены значимые отклонения в проницаемости кишечника для жира и d-ксилозы (в 1,5-3 раза). Там же для 20 пациентов с псориазом выполнена эндоскопия и у всех был установлен хронический диффузный гастродуоденит.

Эти данные означают, что у всех больных псориазом есть функциональные нарушения слизистой кишечника, которые, в частности, серьезно влияют на его проницаемость. Они могут проявляться как из-за наследственной склонности, так и в результате гастроэнтерологических заболеваний.

## 2. Стрептококки в коже и кишечнике.

Стрептококки (род Streptococcus) являются факультативными анаэробами, их классифицируют по наличию специфических углеводов в клеточной стенке и выделяют 17 групп, обозначаемых заглавными латинскими буквами. Независимо используют классификацию Брауна, основанную на особенностях роста стрептококков на агаре с кровью барана. По этой классификации выделяют α-стрептококки (частичный гемолиз и позеленение среды, т.е. зеленящие), β-стрептококки (полностью гемолизирующие или гемолитическими) и γ-стрептококки (дающие визуально невидимый гемолиз).

Возбудителями серьезных болезней человека являются β-стрептококки (BS), большая часть которых относится к группе А. Они являются причиной фарингитов и скарлатины, целлюлитов, рожистых воспалений и стрептодермий [18]. Кожные проявления BS-инфекций сопровождаются гиперемией, образованием фликтен, экссудата и корок. Кожные покровы имеют врожденный иммунитет на стрептококковые инфекции, который состоит из различных механизмов защиты в зависимости от типа BS, степени и длительности поражения кожи. Иммунный ответ сопровождается воспалением, ускоренной пролиферацией кератиноцитов, и, как следствие, образованием корок.

BS-антигены при фокальной инфекции (в первую очередь S. pyogenes) часто приводят появлению на коже временных высыпаний (каплевидный псориаз), а иногда и постоянных (хронический пятнистый псориаз). Это происходит в ситуации, когда собственно инфекционный процесс локализован далеко от наружных покровов кожи. Точный механизм этого явления до конца не изучен, но обзор большого объема данных представлен в [3, глава 5].

Большинство стрептококков группы А (Group A Streptococci - GAS) принадлежит виду S.pyogenes, поэтому оба термина часто используют как синонимы. В работах [3,11,12,13,14] большинство результатов получены для BS принадлежащих группе А.

Стрептококки всегда (как комменсалы) присутствуют в микрофлоре кишечника (их число в фекалиях превышает $10^7$ КОЕ/г)[16]. Чаще обнаруживают α-стрептококки, но и присутствие β-стрептококков (BS) не является причиной каких-либо кишечных заболеваний. В работе [20] сделан анализ 88 случаев заболеваний, вызванных BS (распределение по серогруппам следующее: 43% - А, 27% - В, 4% - С и 26% - G). Там же отмечено, что BS являются нормальной флорой глотки, кожи, кишечного тракта и вагины, но при ослаблении иммунитета (или других причинах) проявляют свою патогенность (ни у одного из обследованных пациентов местом проявления BS-патогенности не был кишечный тракт). О присутствии стрептококков в разных отделах ЖКТ можно судить по данным (Таблица 1) [19].

**Таблица 1. Численность некоторых видов стрептококков в отделах кишечника и фекалиях**

|  | Тощая | Подвздошная | Ободочная | Фекалии |
|---|---|---|---|---|
| a-Streptococcus (S. mutans, S. salivarius и др.) - оральные | 261 | 253 | 1170 | 1691 |
| Streptococcus прочие, кишечные | 1642 | 127 | 2 | 641 |

Примечание. Значения = численность (N х $10^6$ клеток/г)

Статьи [21,22], посвященные статистическому анализу хирургических операций на желудочно-кишечном и желчном тракте, отмечают несомненное присутствие BS в кишечнике. Это присутствие было латентным в дооперационный период и весьма неблагоприятно сказалось на процессе выздоровления в постоперационный период.

Публикация [23] о заболевании ребенка, имевшего трудно диагностируемый перианальный BS-дерматит и, как следствие, каплевидный псориаз, дает основание предположить, что занесение BS на слизистую толстого кишечника может происходить из перианальной области.

Продукты метаболизма и распада микроорганизмов, составляющих биопленку кишечника, попадают в кровоток и при нормальной проницаемости. Их присутствие в крови является постоянным. Этот факт лежит в основе определения процентного состава микрофлоры кишечника по микробным маркерам в крови методом газовой хроматографии в сочетании с масс-спектрометрией (ГХ-МС) [24,19]. Этим методом доказано, что *"около 25% биомассы пристеночной микрофлоры представлено аэробными кокками (стафилококки, стрептококки, энтерококки и коринеформные бактерии)..."* [19].

Авторы работы [36] показали, что у больных псориазом имеет место эндотоксинемия. Был изучен антибактериальный гуморальный иммунитет по отношению к нормальной и условно-патогенной кишечной микрофлоре у больных с псориазом (32 пациента) и у группы здоровых лиц (120 человек). Уровень сывороточных антител определялся к 9 различным группам бактерий, но повышенным он оказался только к S. pyogenes. У больных псориазом он в среднем более чем в два раза превышал норму (29,8 мкг/мл против 13,75). А у половины пациентов он достигал 60 мкг/мл. Традиционное лечение никак не повлияло на данный показатель.[1]

Изложенные факты дают основание утверждать, что BSP и их фрагменты в норме присутствуют в крови практически каждого человека. Степень их присутствия, размер и ассортимент таких белков (и их фрагментов) определяется двумя главными факторами:

--- проницаемостью кишечника для BSP и их фрагментов
--- численностью и распространённостью BS-колоний в кишечнике

Именно эти факторы играют главную роль в новой модели патогенеза псориаза.

## 3. Модель патогенеза псориаза

Здесь (как и выше) и всюду далее BSP это β-Streptococci Proteins - оболочечные и мембранные белки BS, а также их фрагменты - продукты распада BS-колоний. Вышеизложенные факты, анализ известных методик и сопоставление частных случаев дают основания для следующего определения псориаза:

**Псориаз - это эпидермальная гиперпролиферация, которая является ответом кожной иммунной системы на превышение уровня толерантности к BSP-антигенам. Накопление BSP-антигенов в коже происходит вследствие их накопления в крови, что вызвано повышенной проницаемостью стенок кишечника для BSP. Источником BSP являются продукты распада BS-колоний, включенных в микробиоценоз слизистой кишечника. Повышенная проницаемость стенок кишечника для BSP, в первую очередь, обусловлена генетически, но также может быть вызвана и другими, негенетическими факторами.**

Это определение - краткая формулировка взаимодействия нормальных и патологических факторов при псориазе, изображенная на схеме (рис.1). Новым в этой модели патогенеза по сравнению с моделью, предложенной в книге [3], являются псорафакторы 1 и 2, а также другая формулировка псориатического цикла. Остальные факторы с аналогичными взаимосвязями подробно изложены и обоснованы в [3, глава 8]. Прокомментируем каждый из факторов и приведем аргументы в их поддержку.

### Фактор 1. Повышенная проницаемость кишечных стенок для определенных белков (псорафактор-1)

Это нарушение генетически обусловлено и/или приобретено и может иметь место как толстого, так и для тонкого кишечника. Проницаемость повышена, в первую очередь, для BSP-антигенов, продуктов распада BS (β-стрептококков) - белков клеточных оболочек и мембран (а также их фрагментов), которые являются антигенами для кожной иммунной системы. Повышенная проницаемость стенок кишечника имеет место для больных псориазом, что доказано рядом исследований. Лечение именно этого нарушения лежит в основе эффективного метода Пегано [4] (см. раздел 4).

По нашему мнению псориатическая гастроинтестинопатия [1] первична по отношению к кожным проявлениям псориаза, поскольку одним из ее следствий является повышенная проницаемость кишечника для белков. Мы предполагаем, что фактор "повышенной проницаемости кишечника для определенных белков" является генетически наследуемым и его можно назвать "псорафактором-1".

---
[1] Добавлено при переводе

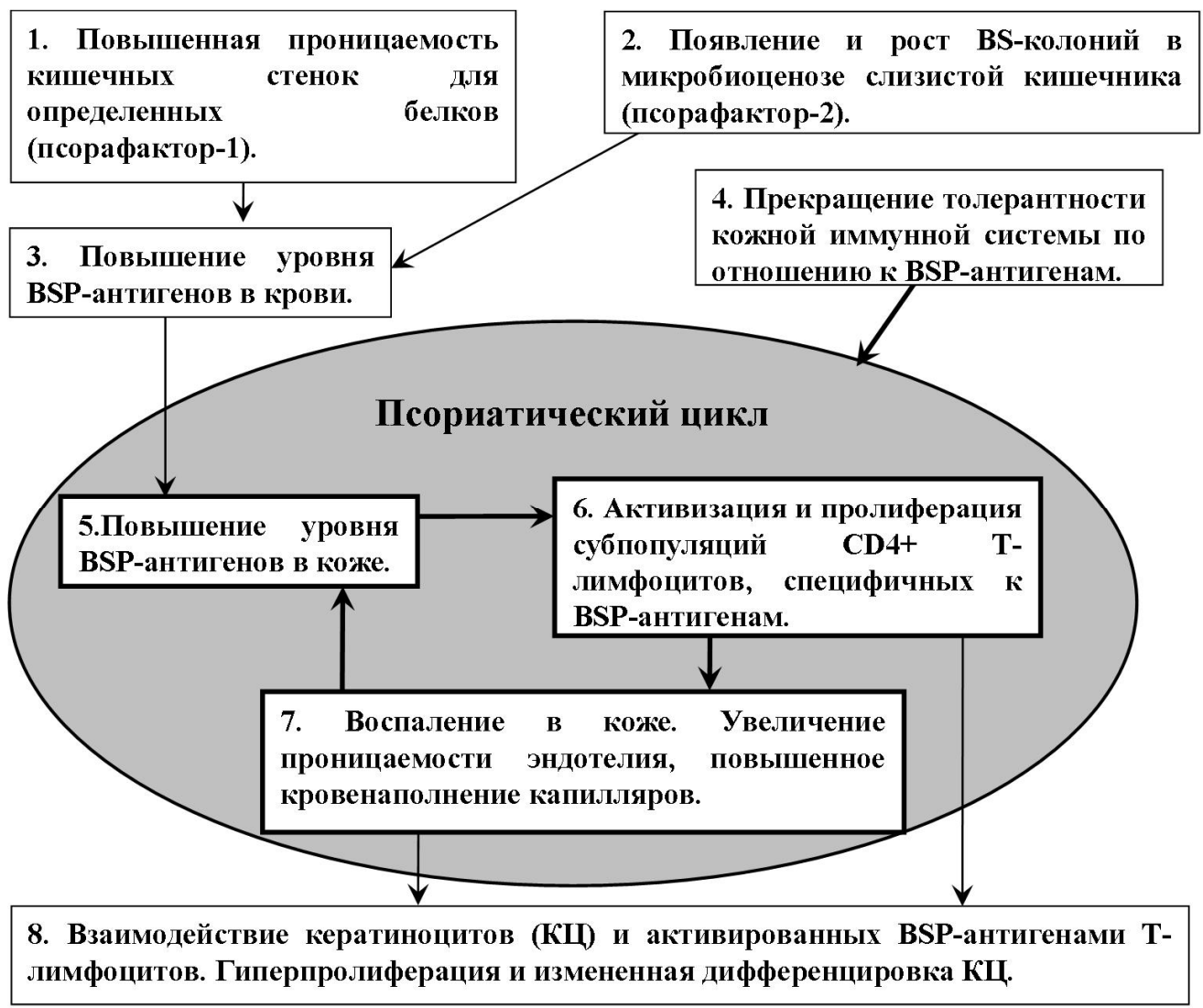

Рис. 1. Модель патогенеза псориаза. Взаимодействие факторов.

Неоспорима корреляция между псориазом и нарушениями функционирования кишечных стенок. Так, например, заболеваемость псориазом при болезни Крона составляет 22% [7], взаимосвязь предполагаемых локусов болезни Крона и псориаза отмечается в [3, раздел 1.2.5].

Гастроэнтерологические заболевания, поражающие кишечные стенки, ухудшают псориаз, т.к. способствуют повышению проницаемости кишечных стенок, что подтверждается взаимосвязью тяжести поражения кишечных стенок с тяжестью псориаза [2].

В 1999 г. в Швеции был проведен эксперимент по лечению псориаза безглютеновой диетой (см. раздел 4). Точный механизм влияния глютена на псориаз неизвестен. Однако известно, что глютен может (при наличии предрасположенности) отрицательно влиять на состояние энтероцитов (что и происходит у больных целиакией), в результате чего атрофируются ворсинки, нарушается всасывание углеводов, белков и жиров [17]. Известно, что при глютеновой энтеропатии имеет место в десятки раз (60 нг/мл при норме 1-4 нг/мл) повышенная проницаемость для белка овальбумина (молекулярная масса 43 kDa) [25]. Можно предположить, что это коррелирует с повышенной проницаемостью кишечных стенок для BSP молекулярной массы 20-50 kDa [13]. Именно таким образом может определяться положительное влияние безглютеновой диеты на псориаз у части пациентов, имеющих склонность к целиакии. Именно так глютен, не приводя к целиакии, может способствовать возникновению и поддержанию псориаза - повышая проницаемость кишечных стенок для BSP.

Так, например, для одного из пациентов дважды (с интервалом в год) был проведен стандартный тест на барьерную функцию тонкой кишки (проницаемость тонкого кишечника) для овальбумина [25]. При наличии псориаза и отсутствии каких-либо серьезных проблем с ЖКТ этот тест показал многократное превышение нормы (14-58 нг/мл при норме 1-4 нг/мл). Вот они открытые ворота для BSP-антигенов…

## Фактор 2. Появление и рост BS-колоний в микробиоценозе слизистой кишечника (псорафактор-2)

BS - это обычные непатогенные обитатели (комменсалы) слизистой кишечника. Объемный рост их колоний происходит за счет увеличения площади и толщины покрытия слизистой толстого кишечника, и, возможного их перехода на слизистую тонкого. Как следствие, растут объемы продуктов их распада, BSP-антигенов.

Именно этот фактор является новым в модели патогенеза псориаза. Предположение о влиянии BS-колоний в микробиоценозе кишечника на развитие псориаза вполне естественно вытекает из моделей патогенеза псориаза, изложенных в [3]. Автор этой книги делает вывод о безусловной причастности BSP-антигенов к развитию не только каплевидного псориаза, но и к развитию и поддержанию хронического пятнистого псориаза. В то время как временный каплевидный псориаз провоцирует временная BS фокальная инфекция (а в частных случаях временный BS перианальный дерматит), то вполне естественен вопрос - где именно в организме человека могут постоянно располагаться BS-колонии не вызывая никаких других последствий кроме поддержания хронического псориаза? Ответ прост - единственное место, где патогенные BS не проявляют свою патогенность - это слизистая кишечника. Они могут существовать здесь в течение всей жизни пациента, не привлекая внимания ни его, ни врачей, поскольку здесь они являются комменсалами.

Заселение слизистой BS-колониями начинается с толстой кишки. Уже при этом, если нарушена проницаемость по белкам (псорафактор-1), может возникнуть псориаз. Это первая стадия развития псориаза – период, когда BS-колоний нет в тонкой кишке. В этот период ситуацию можно полностью обратить, ибо микробиоценоз толстой кишки реально восстановить путем курса гидроколонотерапии с добавками (адсорбенты, урина, яблочный уксус, бифидумбактерин и т.д.). Т.е. полностью уничтожить BS-колонии, устранить одну из первопричин псориаза и, тем самым устранить сам псориаз, по крайней мере, до следующего заселения толстой кишки BS.

Если не принимаются меры к устранению BS из толстой кишки, а моторика и состояние кишечника пациента таковы, что вероятность заброса BS из толстого в тонкий велика (запоры, слабая баугиниева заслонка), то BS начнут заселение тонкой кишки. Эта вторая стадия плоха тем, что устранить BS-колонии из тонкой гораздо сложнее, ибо, во-первых, слизистая тонкой кишки ворсинчатая, а, во-вторых, колонотерапия не захватывает тонкую кишку. Поверхность тонкой (включая ворсинки) значительно выше, чем у толстой – как следствие и псориаз может развиться масштабнее.

При этом плохо также то, что полная очистка толстой кишки от BS становится недостижима, ибо постоянно будет происходить подзаселение BS, приходящих из тонкой кишки вместе с химусом.

Из вышесказанного ясно как важно начинать лечить псориаз как можно скорее после его начала! Становятся понятны успешные результаты тех, кто недавно заболел, поскольку такие пациенты потенциально могут вылечиться полностью[2]. И понятны трудности тех, кто начал лечение, когда первопричина псориаза - заселение слизистой кишечника BS, перешла во вторую стадию – стадию захвата тонкой кишки (по-видимому, таких пациентов большинство).

Также понятны успехи тех пациентов, которые применяют специальные диеты и лечебное голодание. Помимо восстановления кишечных стенок и возможно частичного снижения проницаемости по белкам, такие методики, безусловно, влияют на кишечную микрофлору. Они наносят удар по микробиоценозу кишечника в целом и, в частности, по BS-колониям. Длительное сухое голодание [26] может привести к полной элиминации отдельных видов микроорганизмов и, в частности, BS-колоний. Именно этим, по нашему мнению, объясняется успех такого рода лечения псориаза.

При переходе на режим Пегано иногда очищению кожи предшествует временное ухудшение, так называемая реакция Герксхаймера [4]. Она хорошо известна врачам, поскольку часто проявляется в начале антибиотикотерапии многих бактериальных инфекций. Причиной этой реакции является массовая гибель микроорганизмов, сопровождающаяся выделением продуктов их распада. Ухудшение кожных проявлений, в частности, определяется именно резким ростом объемов продуктов распада. Явление это временное, преходящее и обычно не требует никакого специального лечения. Оно проходит по мере сокращения объемов колоний патогенных микроорганизмов, которые гибнут под воздействием антибиотиков.

Следует отметить, что в редких случаях реакция Герксхаймера может привести к более серьезному, чем кожные высыпания, эффекту, так называемому токсическому шоку, что потребует медикаментозного вмешательства.

Ухудшение кожных высыпаний в начале перехода на режим Пегано, как можно предположить, возникает из-за той же причины - массовой гибели BS-колоний, существующих на слизистой кишечника (см. раздел 4). Переход на данный режим меняет условия, в которых прижились и размножились BS-колонии, и, как следствие, они начинают активно гибнуть из-за недостатка необходимых для них питательных веществ. Результат - продукты их распада попадают в кровь, достигают кожи, а кожная иммунная система получает временный дополнительный стимул для активизации своего ответа.

Тот факт, что переход на режим Пегано не всегда сопровождается этой реакцией, означает, что либо а) постепенность перехода на режим определяет постепенность гибели BS-колоний, либо б) благодаря очистительным процедурам живые BS-колонии отрываются от слизистой кишечника и выводятся вместе с химусом и фекалиями из организма, либо в) нарушенная проницаемость кишечных стенок восстанавливается опережающими темпами по сравнению со скоростью гибели BS-колоний.

Последнее вполне возможно, так как диета Пегано ограничивает потребление животных белков, что снижает нагрузку на механизмы расщепления, транспортировки и всасывания белков в кишечнике. Предположим, что повышенная проницаемость для BSP-антигенов связана в том числе и с перегрузкой этих механизмов. Тогда уменьшение такой

---

[2] Отзывы на книгу Дж.Пегано, www.psora.df.ru/responses.html, респонденты Alexvk, Надя, Glenn

нагрузки одновременно повышает качество работы этих механизмов, и, как следствие, снижает проницаемость кишечных стенок для BSP-антигенов.

Применение энтеросорбентов дает хороший результат при псориазе (см. раздел 4). По нашему мнению их эффективность определяется тем насколько высокий процент продуктов распада BS-колоний адсорбируется (и затем выводится через кишечник из организма) до того, как происходит их расщепление до BSP-антигенов и последующий транспорт через сверхпроницаемые кишечные стенки в кровь.

Грубые методики (применение антибиотиков, длительное голодание и др.) одновременно с воздействием на BS-колонии могут так нарушить микробиоценоз в целом, что это приведет к дисбактериозу. Ясно, что оптимальной методикой была бы та, которая обеспечивала бы селективную элиминацию BS-колоний. Это можно было бы обеспечить путем подбора индивидуального BS-фаголизата и последующего курса лечения, включающего прием per os и per rectum.

Отметим также корреляцию между следующими фактами: а) с хроническим псориазом не рождаются, но он может начаться в любом возрасте и в подавляющем большинстве случаев сохранится до конца жизни пациента; б) человек рождается со стерильным кишечным трактом, становление микробиоценоза кишечника происходит постепенно по мере взросления, изменения микробиоценоза могут происходить в любом возрасте и, при отсутствии специального лечения или значимой смены диеты, сохраняются на всю жизнь.

Также отметим известную взаимосвязь между более ранним проявлением псориаза и искусственным вскармливанием младенцев. Известно, что искусственное вскармливание чаще приводит к дисбалансу кишечного микробиоценоза, снижает колонизационную резистентность. В рамках данной модели ясно, что именно в таких условиях BS-колонии имеют больше шансов закрепиться и вырасти на слизистой кишечника в раннем возрасте пациента.

В рамках данной модели можно предположить, что семейная предрасположенность к псориазу помимо генетической компоненты имеет также и бытовые причины. Ребенок (подросток), имея генетически обусловленную повышенную проницаемость кишечника к белкам, и находясь рядом (постоянно или периодически) с близким родственником, имеющим псориаз, с большей вероятностью вступит в контакт с BS, носителем которого является этот родственник. Да, псориаз незаразен, но имеющий псориаз, имеет в микробиоценозе своего кишечника BS (псорафактор-2), опасный для тех, кто имеет повышенную проницаемость кишечника для BSP-антигенов (псорафактор-1).

### Фактор 3. Повышение уровня BSP-антигенов в крови

Это происходит благодаря сочетанию псорафакторов 1 и 2.

Данный фактор хорошо известен при каплевидном псориазе, а также имеет место и при хроническом пятнистом (бляшечном) [3]. В отличие от каплевидного псориаза, когда причиной повышения уровня BSP-антигенов является BS-фокальная инфекция, в случае хронического пятнистого псориаза источником BSP-антигенов являются продукты распада BS-колоний вошедших в состав кишечного микробиоценоза на постоянной основе. Продукты метаболизма и распада микроорганизмов попадают в кровоток всегда (см. раздел 2), но именно повышенная проницаемость кишечника для определенных белков (псорафактор-1) приводит к чрезмерному росту уровня BSP-антигенов. Как минимум, он достигает тех же значений, что и при BS-фокальной инфекции.

Кардинальное снижение этого уровня, по нашему мнению, обеспечивает такая небезопасная методика лечения тяжелого псориаза как гемосорбция [27]. При гемосорбции из крови фильтруются циркулирующие иммунные комплексы (ЦИК) и среднемолекулярные белки. Однако с какими именно антигенами сцеплены антитела в ЦИК, точно не установлено, также как неясен перечень фильтруемых среднемолекулярных белков. Во многих публикациях по псориазу неопознанный элемент, фильтруемый при гемосорбции, называется "псориазобразующим началом" [28].

В рамках данной модели логично предположить, что при гемосорбции, в первую очередь, фильтруются BSP-антигены, что и приводит, в конечном счете, к ремиссии псориаза. BSP-антигены и являются "псориазобразующим началом".

Фильтрация ЦИК при гемосорбции дает второй по значимости эффект, ибо снижает нагрузку с печени, в чьи функции входит фильтрация, утилизация и детоксикация. Снижение уровня ЦИК в крови также останавливает процесс их отложения на эндотелиальных стенках микрососудов, что в свою очередь восстанавливает их проницаемость [29].

На поддержку работы печени направлены многие очистительные процедуры (например, дюбаж) и препараты (например, Эссенциале, Эссливер Форте), которые приводят к ее более эффективной работе по утилизации BSP-антигенов. Эти процедуры и препараты активизируют работу печени так, что она становится способна очищать кровь от избытка BSP-антигенов. В результате достигается хороший результат на коже, но перегружается печень, что, в конечном счете, может сказаться на здоровье пациента. Кроме того, при этом не устраняется причина псориаза, а только временно снижается уровень BSP-антигенов в крови.

Аналогичным образом действует гемодез - водно-солевой раствор поливинилпирролидона, ионов калия, натрия, магния, хлора. Введенный внутривенно он в короткий срок (до 80% выводится за 4 часа) связывает токсины и выводит их через почечный барьер [30]. Эффективность гемодеза при тяжелом псориазе и псориатическом артрите общеизвестна. Однако какое именно "псориазобразующее начало" он связывает неизвестно. В рамках данной модели логично предположить, что это BSP-антигены.

### Фактор 4. Прекращение толерантности кожной иммунной системы по отношению к BSP-антигенам

Происходит из-за превышения порогового уровня и/или благодаря костимуляции[3], поскольку происходит одно из пусковых (триггерных) событий:

- BS фокальная инфекция
- Стрептодермия
- Инфекционное заболевание
- Травма эпидермиса (эффект Кебнера)
- Эндокринные отклонения
- Прием определенных медикаментов
- Стресс
- Алкоголь

---

[3] Костимуляцией Т-лимфоцитов называется процесс, когда в дополнение к сигналу, индуцированному через их ТкР (Т-клеточный рецептор), они одновременно получают второй сигнал через свой другой клеточный рецептор CD28. Часто только при этом увеличивается секреция цитокина ИЛ-2, и, как следствие, происходит активизация и пролиферация Т-лимфоцитов. Отсутствие костимуляции может привести к игнорированию антигенов, либо к развитию анергии. [29, стр. 265-6]

Повышение уровня BSP-антигенов в коже происходит постепенно, по мере объемного роста BS-колоний на слизистой кишечника и повышения проницаемости кишечных стенок по отношению к белкам.

Реакция кожной иммунной системы на BSP-антигены, как и на другие BS-антигены естественна. Она обеспечивает ускоренное обновление кожи (гиперпролиферацию), предполагая стрептодермию (см. раздел 2). При стрептодермии ускоренное воспроизводство КЦ обеспечивает ускоренное сбрасывание (слущивание) пораженных слоев кожи, что является простейшей формой защиты от инфекции.

Начало реакции кожной иммунной системы происходит либо при превышении порогового уровня BSP-антигенов, либо при наступлении пускового события. В результате специфические к BSP-антигенам Т-лимфоциты выходят из состоянии анергии и активизируются.

Отметим, что любое из пусковых событий одновременно является событием, способствующим завершению ремиссии и началу рецидива у пациента со стажем.

Различные пусковые события по-разному обеспечивают начало (или усиление) реакции кожной иммунной системы. Так, травма сама вызывает воспалительный процесс, что увеличивает проницаемость эндотелия, т.е. напрямую влияет на фактор 7. Алкоголь воздействует как на кишечные стенки, увеличивая их проницаемость (фактор 1), так и на кровенаполнение и проницаемость периферийных капилляров (фактор 7), тем самым является двойным стимулятором. BS-фокальная инфекция и стрептодермия способствуют увеличению BSP-антигенов в крови (фактор 3). Известно, что эндокринные отклонения могут влиять на интенсивность и тип иммунного ответа (фактор 6). Этим же обусловлено влияние стресса, ибо он активно влияет на эндокринное состояние организма.

Инфекционные заболевания способствуют увеличению уровня активированных (не BSP-антигенами) Т-лимфоцитов в крови и коже, что приводит к активизации антигенпрезентирующих клеток Лангерганса (LC). Они в свою очередь экспрессируют рецепторы B7 на своей поверхности, что обеспечивает костимулирующий сигнал при презентации BSP-антигенов. В конечном счете это, через воздействие цитокинов, способствует активизации специфичных к BSP-антигенам Т-лимфоцитов в коже и/или в региональных лимфоузлах (фактор 6).

## Фактор 5. Повышение уровня BSP-антигенов в коже

Это происходит благодаря их миграции через эндотелий, также как и при BS-фокальной инфекции [3].

Кожа является одним из элиминативных органов, который дополняет работу печени и почек по очистке крови от токсинов. Постоянное слущивание КЦ обеспечивает постоянное удаление из организма токсинов, а в экстренных случаях (например, кишечное отравление или инфекционное заболевание) возникают различного рода временные высыпания. Хронический псориаз, по сути, является постоянным ускоренным освобождением крови от BSP-антигенов (фактор 3).

Известна особенность стационарных ("дежурных") псориатических бляшек в большей степени проявляться на локтях, коленях, ступнях и кистях. Процесс попадания BSP-антигенов из артерий в микроциркуляторную кожную систему взаимосвязан со структурой артерий, скоростью прохождения через них крови, наличием турбулентных завихрений. Перечисленные выше места весьма характерны изгибами и разветвлениями артерий, что способствует большей концентрации BSP-антигенов в микроциркуляторной кожной системе и, как следствие, их большей инфильтрации в кожу.

Положительное влияние водных процедур, таких как баня, моржевание, паровые ванны, контрастный душ, основано, в том числе, на активной стимуляции всей микроциркуляторной кожной системы. Ее стимуляция способствует более равномерной (по всей поверхности) инфильтрации BSP-антигенов в кожу и влечет сокращение бляшкообразования.

### Фактор 6. Активизация и пролиферация субпопуляции CD4+ Т-лимфоцитов, специфичных к BSP-антигенам

Это происходит благодаря сочетанию факторов 4 и 5. BSP-антигены поглощаются клетками Лангерганса (LC) и процессируются. Затем LC презентирует процессированные BSP-антигены присутствующим в эпидермисе антиген-специфичным CD4+ Т-лимфоцитам (Т-хелперам), одновременно другие LC доставляют их через афферентный лимфососуд в региональный лимфоузел, где они презентируются аналогичным Т-лимфоцитам. В условиях, созданных конкретным пусковым событием, Т-лимфоциты активизируются и пролиферируют.

Активизация эпидермальных Т-лимфоцитов приводит к выделению цитокинов, которые стимулируют экспрессию адгезивных молекул эндотелиальными клетками постартериальных венул.

Пролиферация Т-лимфоцитов в региональном лимфоузле приводит к образованию новых активированных Т-лимфоцитов (в т.ч. и Т-лимфоцитов памяти), которые попадают через выносящий лимфатический сосуд в кровоток. Экспрессия на них хоминг-рецепторов обеспечивает их миграцию в кровотоке к постартериальным венулам, на которых экспрессированы адгезивные молекулы, т.е. именно туда, где LC активировали эпидермальные Т-лимфоциты.

Именно на эндотелии постартериальных венул благодаря взаимодействию их рецепторов и адгезивных молекул происходит их инфильтрация из кровотока в дерму, а затем под действием хемокинов в эпидермис. Увеличение числа активированных CD4+ Т-лимфоцитов (Т-хелперов) в конкретном месте эпидермиса означает еще большее выделение цитокинов, в частности, обеспечивающих увеличение проницаемости эндотелия для лейкоцитов (нейтрофилов, макрофагов и лимфоцитов). В месте будущего псориатического высыпания начинается воспалительный процесс. То, что псориатический процесс связан с увеличением числа Т-хелперов в эпидермисе хорошо известно [8,30].

Известно, что клетки Лангерганса (LC) очень чувствительны к воздействию ультрафиолета (УФ) типа В и под его воздействием инактивируются [29,30]. В результате презентация BSP-антигенов приостанавливается, активизация CD4+ Т-лимфоцитов также, и гиперпролиферация КЦ замедляется. Именно на этом основаны гелио- и УФ-терапии псориаза [4,30], приводящие к временной ремиссии. В первую очередь именно этим в большинстве случаев определяется сезонный характер псориаза: летом световой день дольше, погода теплее - одежды меньше и она более прозрачна, в результате кожа получает больше УФ - клетки Лангерганса инактивируются. С наступлением осени и зимы ситуация меняется, кожа получает все меньше и меньше УФ - клетки Лангерганса активируются, наступает рецидив псориаза.

Следует отметить, что эти и другие традиционные терапии [30], подавляющие кожный иммунный ответ, не могут привести к положительному постоянному результату, т.е. полному освобождению от псориаза, так как не влияют на псорафакторы. Более того, во время подавления кожного иммунного ответа ускоренное слущивание приостанавливается, а фактор 3 продолжает действовать, в результате уровень BSP-антигенов в коже повышается выше обычного. И, когда действие такой терапии прекращается, часто возникает рецидив, интенсивность которого определяется этим повышенным уровнем.

Огромный практический опыт привел профессора Потекаева Н.С. к выводу о том, что у пациентов с псориазом никогда не бывает пиодермии (частное сообщение). По нашему мнению стрептодермию предотвращает постоянный кожный иммунный ответ на мнимое присутствие BS. Таким образом наличие BSP-антигенов в коже обеспечивает своеобразную постоянную вакцинацию против стрептодермий.

### Фактор 7. Воспаление в коже. Увеличение проницаемости эндотелия, повышенное кровенаполнение капилляров

"Воспаление - это реакция организма, обеспечивающая привлечение лейкоцитов и растворимых компонентов плазмы в очаги инфекции или повреждения ткани. К его основным проявлениям относятся повышение кровенаполнения капилляров и их проницаемости для сывороточных макромолекул, а также усиленная миграция лейкоцитов в направлении воспалительного очага через эндотелий расположенных поблизости сосудов." [29, стр.83]

Увеличение проницаемости эндотелия для сывороточных макромолекул влечет увеличение его проницаемости и для BSP-антигенов. Повышенное кровенаполнение капилляров способствует локальному увеличению числа BSP-антигенов и, как следствие, повышение их уровня в коже.

Необходимо отметить роль ЦИК, отложение которых на стенках эндотелия, повышает их проницаемость [29, глава 25]. Увеличенное число ЦИК в крови при псориазе является следствием гуморального ответа на присутствие в крови как BSP-антигенов, так и любых других антигенов (вот еще почему инфекция, вызывающая гуморальный иммунный ответ, также является пусковым фактором) [27,32]. Места отложений ЦИК могут определять расположение высыпаний.

Факторы 5, 6 и 7 представляют собой псориатический цикл для хронического пятнистого псориаза (на схеме включены в овал). На фоне действия фактора 3, этот цикл запускается одним из костимулирующих событий (фактор 4), но продолжает функционировать и после завершения действия этого фактора, поскольку начавшийся воспалительный процесс в коже (фактор 7) увеличивает проницаемость эндотелия, в том числе и для BSP-антигенов. В результате, воспаление вместо того, чтобы приводить к устранению своей причины, наоборот, приводит к ее стимуляции. Т.е. после начала функционирования этого цикла для его поддержки достаточно действия одного фактора 3.

Традиционные способы лечения псориаза разрывают этот цикл, действуя на факторы 6 или 7, одновременно, в той или иной степени, устраняя (подавляя) костимулирующие факторы 4. Однако в полной мере результативным будет лечение, направленное на устранение (подавление) хотя бы одной из причин фактора 3, т.е. на псорафакторы 1 или 2.

Только полное устранение BSP-антигенов из крови приведет, в конечном счете, к их устранению из кожи и к прекращению воспалительного процесса.

В случае каплевидного псориаза BS-фокальная инфекция приводит к временному присутствию BSP-антигенов в крови. Поэтому, по мере ее завершения, они устраняются из крови, успев, однако, явиться причиной временных каплевидных псориатических высыпаний.

**Фактор 8. Взаимодействие кератиноцитов (КЦ) и активированных BSP-антигенами Т-лимфоцитов. Гиперпролиферация и измененная дифференцировка КЦ**

Взаимодействие КЦ и активированных Т-лимфоцитов происходит после образования молекул адгезии на КЦ, чему способствуют цитокины, продуцируемые благодаря фактору 6. Взаимодействие идет через молекулы адгезии, другие поверхностные молекулы и через местную продукцию цитокинов.

Гомология между BSP-антигенами и кератинами (поверхностными белками КЦ) из-за перекрестной реактивности может играть стимулирующую роль, в частности, для фактора 6, подталкивая начавшийся псориатический цикл.

Этот процесс является частью хронического воспаления, поскольку ускоренное слущивание эпидермальных клеток является попыткой кожной иммунной системы решить проблему растущего присутствия в коже BSP-антигенов. По-видимому, таким образом, кожная иммунная система, исчерпав другие возможности, пытается избавиться от мнимого присутствия BS в коже. Ускоренное слущивание является действенным механизмом при многих кожных инфекциях, когда инфицирующие кожу бактерии в ней и локализованы.

В случае с хроническим псориазом мы имеем ситуацию, когда кожная иммунная система срабатывает на увеличенное присутствие BSP-антигенов, ошибочно предполагая (как при стрептодермии) присутствие BS в коже. Она последовательно включает все необходимые воспалительные механизмы, в т.ч. (и, наверное, не сразу) ускоренное слущивание, но именно один из этих механизмов, увеличение проницаемости эндотелия, приводит к еще большему увеличению присутствия BSP-антигенов в коже. Тем самым, воспалительный процесс, парадоксально, вместо устранения спровоцировавшей его причины приводит к ее усилению.

Так начинается и поддерживается гиперпролиферация и измененная дифференцировка КЦ, т.е. хронический псориаз.

## 4. Практика лечения

Метод Пегано [4] был разработан более 20 лет назад в США и с тех пор получил широкое распространение по всему миру, в том числе и в России. Причина псориаза по Пегано: "нарушение барьерной функции кишечника и проникновение токсинов в кровеносную и лимфатическую системы". Пегано не конкретизирует для чего именно нарушена барьерная функция и какие именно токсины вредны при псориазе, кроме того он никак не учитывает роль кожной иммунной системы. Однако его метод в целом направлен на устранение псорафактора 1 и поэтому очень часто дает положительный результат.

Режим Пегано состоит из диеты, регулярного очищения кишечника, фиточаев и терапии позвоночника. Очищение кишечника обеспечивается колонотерапией, клизмами, а также разгрузочными фруктовыми диетами. Важная роль отводится потреблению чистой воды (1,2-1,6 литра в сутки). Диета Пегано содержит ряд строгих ограничений, таких как запрещение табакокурения и алкоголя, исключение острых, жареных, жирных и соленых продуктов, исключение мяса (кроме баранины и птицы), исключение ракообразных, моллюсков и т. д.

**Таблица 2. Проницаемость кишечника, критерии PASI и PSS до начала режима Пегано и после шести месяцев его соблюдения.**

|  | Отношение лактулоза/маннитол | | PASI | | PSS | |
|---|---|---|---|---|---|---|
| Случай | До | После | До | После | До | После |
| 1 | 0,134* | 0,038 | 7,0 | 4,8 | 7,0 | 6,0 |
| 2 | 0,084* | 0,022 | 30,7 | 18,4 | 14,0 | 5,0 |
| 3 | 0,034 | 0,019 | 14,0 | 0,7 | 21,0 | 3,0 |
| 4 | 0,047 | 0,024 | 2,3 | 0,0 | 7,0 | 1,0 |
| 5 | 0,029 | 0,026 | 37,0 | 19,8 | 24,0 | 12,0 |
| В среднем | 0,066* | 0,026 | 18,2 | 8,7 | 14,6 | 5,4 |

\* Вне нормы, границы которой составляют 0,01–0,06.

В Таблице 2 отображены результаты эксперимента, проведенного в США в 1999 г. [6]. В нем участвовали 5 пациентов с разной степенью выраженности псориаза. До начала соблюдения режима Пегано у всех наблюдалось отклонение относительной проницаемости кишечника (лактулоза/маннитол) от нормы. Этот параметр пришел в норму у всех после 6-ти месяцев соблюдения режима. Состояние кожи пациентов при этом также улучшилось (см. PASI). Результаты успешного применения метода Пегано есть и в России, о чем свидетельствуют положительные отзывы дерматологов и пациентов.

Другой метод был применен при лечении 30 глютензависимых пациентов с псориазом. Вначале в 1997-2000 г. в госпитале в Уппсала (Швеция) были обследованы более 300 пациентов с псориазом [33]. У 16% из них было выявлено наличие антител к глиадину. Было высказано предположение, что изменение диеты этих пациентов может способствовать улучшению состояния их кожи. Чтобы подтвердить эту гипотезу в 1999 г. был проведен эксперимент для 30 таких пациентов. В течение 3-х месяцев они соблюдали безглютеновую диету. Прием всех прочих лекарств и процедур был сохранен. За три месяца улучшение наступило у 73% пациентов, у 10% состояние осталось без изменений, у 17% - ухудшилось. Индекс PASI в среднем уменьшился с 5,5 до 3,6. По нашему мнению безглютеновая диета обеспечила частичное восстановление барьерной функции кишечника по отношению к BSP-антигенам.

Известен положительный эффект от применения энтеросорбентов. Так в работе [8] описано примение препарата "Силлард" (в России выпускается как "Полисорб"), состоящий из высокодисперсного кремнезема. Лечение начинают с 10–14-дневного курса терапии силлардом в дозе 1 г три раза в день за 1 час до еды или через 1,5 часа после еды. Улучшение состояния наступает в первые дни лечения и не сопровождается осложнениями. В работе [34] описан результат такой терапии (с включением фитосборов) для 50 пациентов. Клиническое выздоровление наступило у 28 больных, значительное улучшение - у 18, улучшение - у 9. В работе [2, глава 1] содержится обзор результатов успешного применения энтеросорбентов СКНП-1 и СКНП-2 при лечении псориаза.

Эффективность энтеросорбентотерапии по нашему мнению определяется степенью адсорбции продуктов распада BS-колоний (в т.ч. BSP-антигенов) до их всасывания через кишечные стенки в кровь.

В работе [35], выполненной в Венгрии, авторы проверили гипотезу о том, что дефицит желчных кислот (ЖК) играет роль в патогенезе псориаза[4]. Тестовая группа из 551 пациентов с псориазом (до начала эксперимента среднее PASI = 19,1) принимала орально желчную дегидрохолевую кислоту (ДК). По завершении лечения 434 пациента (78.8 %) стали бессимптомными, у остальных 117 наблюдалось значительное улучшение. Среднее PASI после курса лечения стало равным 2.7. Одновременно контрольная группа из 249 пациентов с псориазом получала обычную терапию, по завершении которой только 62 пациента (24.9 %) показали стали бессимптомными. Двумя годами позже 319 из 551 (57.9 %) пациентов тестовой группы оставались бессимптомными, по сравнению только с 15 из 249 (6%) пациентов контрольной группы. Прием ДК приводит к временному увеличению объема ЖК (включая саму ДК), что способствует сокращению объема эндотоксинов (продуктов бактериального распада), транслоцирующихся через кишечные стенки в кровь. Полученные результаты основаны на разрушительной способности ЖК по отношению к эндотоксинам (каким именно не указано).

---
[4] Последние два абзаца статьи добавлены при переводе на английский.

Предположим, что этими эндотоксинами являются BSP-антигены. Однако их разрушения было бы недостаточно для длительной последующей ремиссии. В этом случае по окончании курса лечения должен наступить рецидив, поскольку производящие эндотоксин BS-колонии сохранились. Достигнутая для большинства пациентов длительная бессимптомная ремиссия дает основание предполагать, что избыток ЖК уничтожает BS-колонии. Возможно это происходит путем разрушения ЖК полисахаридных адгезивных связей BS-колоний с мукозно-эпителиальной поверхностью слизистой кишечника [16]. Поскольку избыток ЖК также стимулирует перистальтику, то после разрушения связей происходит отрыв BS-колоний от слизистой и их последующее выведение вместе с химусом из тонкого кишечника и с фекалиями из организма. При этом ремиссия может быть длительной и устойчивой, а рецидив произойдет только при новой BS-колонизации кишечника.

## *Литература*

<u>Примечание</u>. Рефераты (и их переводы) статей, для которых после выходных данных указан RN, находятся на странице www.psora.df.ru/referats.html, где N - порядковый номер.